\documentstyle [12pt] {article}
\begin{document}
\title {Motion in Bi-metric Type Theories of Gravity}
\maketitle
\begin{center}
{\bf { Magd E.Kahil }{\footnote{October University for Modern sciences and Arts , Giza , Egypt}} {\footnote{The American  in Cairo, New Cairo, Egypt \\
e.mail: kahil@aucegypt.edu}}} \footnote{Egyptian Relativity Group. Cairo, Egypt}
\end{center}
\begin{abstract}
{The problem of motion for different test particles , charged and spinning objects of constant spinning tensor in different versions of  the bi-metric theory of gravity is obtained by deriving their corresponding path and path deviation equations, using a modified Bazanski  in the  presence of Riemannian geometry. This method enables us to find the path and path deviation equations of different objects orbiting very strong gravitational fields.}
\end{abstract}

\section{Bi-metric Theories: A Brief Introduction}
General relativity is considered a landmark in history of science of being, during the last century, as a pivotal non-linear theory of gravity [1]. Yet, some problems have been remained  unsolved due to applying Riemannian geometry in its explanation. One of these difficulties is related to the law of conservation of energy and momentum [2], which is dealt  with considering the metrical tensor as flat one at great distances from the gravitational source.
 Rosen [3,4] introduced a remedy to this problem  by proposing two different metrics $g_{\mu \nu}$  representing the gravitational source and giving a curved space and $\gamma_{\mu \nu}$ describing a physical one   expressing an inertial frame and becoming a flat space. Using this assumption, it can be found that the field equations of Einstein imply to a theory of gravitation in flat space . A problem arises because the pseudo-tensor quantities in Orthodox General Relativity,  then turns not to be  preserve its tensor character  [2].  This led Yalmoz [5] to examine a new class of solutions for the field equations of Rosen's theory of gravitation  to solve the difficulty of dealing with the flat metric.  But, a slight problem has emerged in that, the speed of light is no longer constant as confirmed experimentally in the realm of Special Relativity.
Such a problem is counted to be a virtue  using Moffat's approach of a bi-metric theory of gravity for a variable speed of light, as it helps to reveal the puzzle of the dark energy  problem. This approach may be expressed by means of  two metrics are likened to each other in terms of gradients of scalar /biscalar fields to  explain  the rapid expansion of galaxies is due to change of speed of light from one epoch to another that i.e. dark energy can be interpreted  This type of theories are expressing how the inflation scenario of the universe is due to Bi-metric version of variable speed of light [7] . Some applications using the Moffat formalism of bi-metric theory of gravity are explaining the causal description of quantum entanglement [8] and testing the propagation of neutrinos using OPERA experiment [9].

   Another problem in GR is inability to explain the rotation curves of spiral galaxies. Milgram [10] proposed a specific treatment by performing a  modified Newtonian Dynamics paradigm (MOND)that was able to explain its causality  apart from appealing to the dark matter problem. A new step has been taken successively to extend MOND to be expressed in terms of bi-metric theory to become  BIMOND [11] by having two field equations describing  matter and twin-matter[12] may be used for examining the existence of gravitational waves and explaining two interacting 4D membranes[13].
Recently, Hassan-Rosen [14] developed  an extension of the present bi-metric theory using the concept of bi-gravity, which used  two metrics describing their gravitational fields. In doing so, they discard the earlier  bi-metric theories metrics in which  one metric describes gravity while the other is physical. Accordingly, a new massive gravity theory for spin-2 , free from ghosts, has been obtained [15].   Several applications of these types of theories  are viable in dealing with obtaining field equations for very strong gravitational fields of neutron stars [16], which  suggests the possibility  to examine black holes and super-massive black holes e.g Sgr A* by studying the stability of objects orbiting in these fields. This hypothesis is essentially to be examined.

\section{Motion in Bi-metric Theories}
Theories of gravity may help to explain the functioning of different objects. For example,  Rosen[3] obtained the equation of motion   bi-metric theory of gravitation for a test particle. These equations  were  solved by Isrealit[17]  in order to examine their behavior in the case of small velocities and weak fields using a post Newtonian approximation  .

Also, Falik and Opher [18] used  the bi-metric theory of gravity to  find the field equations associated with spinning neutron stars as an example of a strong gravitational field . This achievement opens the way to examine the motion of different charged objects and spinning ones in the presence of a strong gravitational field as defined by Bi-metric type theories. This  led us to obtain the corresponding path and path deviation of different objects , such as test particles, charged particle, spinning objects. These results are determined by  introducing  a Lagrangian  with a specific feature for obtaining the path and path deviations.  Therefore, the key role to  this approach is based to obtain path and path deviation equations for each of these objects using a specified  Lagrangian  for each case. The aim of our study is  obtaining equations of motion for objects using bi-metric theory originated  from two metrics in one stands for gravitational field and the other defines physical matter or both represent gravity.

 \subsection{Path and Path Deviation Equations: The Bazanski Approach }
 Geodesic and geodesic deviation equations can be obtained from the following Bazanski Lagrangian [19]:

\begin{equation}
L = g_{\alpha \beta} U^{\alpha}\frac{D \Psi^{\beta}}{Ds},
\end{equation}
where $g_{\mu \nu}$ is the metric tensor $U^{\alpha}$ is a unit tangent vector to the geodesic , $\Psi^{\beta}$ its deviation vector, $S$ is the parameter characterizing the geodesic and geodesic deviation $\frac{D}{DS}$ is a covariant derivative with respect to $g_{\mu \nu}$. If one takes the variation with respect to $ \Psi^{\rho}$ , we get
\begin{equation}
\frac{\partial L}{\partial \Psi^{\sigma}} = g_{\alpha \beta} \Gamma^{\beta}_{\mu \nu} \delta^{\nu}_{\sigma}U^{\mu},
\end{equation}
and
\begin{equation}
\frac{\partial L}{\partial \dot{\Psi}^{\sigma}} = g_{\alpha \beta} \delta^{\beta}_{\sigma}U^{\alpha},
\end{equation}
\begin{equation}
\frac{d}{dS}\frac{\partial L}{\partial \dot{\Psi}^{\sigma}}= g_{\alpha \sigma , \rho} U^{\alpha}U^{\rho} + g_{\alpha \sigma} \frac{d U^{\alpha}}{dS}.
\end{equation}
Substituting from the above relations into the Euler-Lagrange equation:
\begin{equation}
\frac{d}{dS}\frac{\partial L}{\partial \dot{\Psi}^{\sigma}}- \frac{\partial L}{\partial \Psi^{\sigma}} =0,
\end{equation}
and taking into account the following condition:
\begin{equation}
g_{\mu \nu ; \rho} = 0.
\end{equation}
We obtain after some manipulations:
\begin{equation}
\frac{dU^{\alpha}}{dS} + \Gamma^{\alpha}_{\mu \nu}U^{\mu}U^{\nu}=0,
\end{equation}
where $\Gamma^{\alpha}_{\mu \nu}$ is the Levi-Civita affine connection.
Also, following the same technique can be applied to obtain the variation with respect  $U^{\rho}$ to obtain the geodesic deviation equations:
\begin{equation}
\frac{\partial L}{\partial U^{\sigma}} = g_{\mu \nu} \delta^{\mu}_{\sigma} \frac{D \Psi^{\nu}}{DS}+ g_{\mu \nu} \Gamma^{\nu}_{\alpha \beta} \delta^{\beta}_{\sigma}U^{\mu},
\end{equation}

\begin{equation}
\frac{d}{d S}\frac{\partial L}{\partial U^{\sigma}} = (g_{\mu \nu} \delta^{\mu}_{\sigma} \frac{D \Psi^{\nu}}{DS})_{, \rho}U^{\rho}+ ( g_{\mu \nu} \Gamma^{\nu}_{\alpha \beta} \delta^{\beta}_{\sigma}U^{\mu})_{\rho} U^{\rho},
\end{equation}
and
\begin{equation}
\frac{\partial L}{\partial x^{\sigma}} = g_{\mu \nu , \sigma} U^{\mu} \frac{D \Psi^{\nu}}{DS}+ g_{\mu \nu} \Gamma^{\nu}_{\alpha \beta ,\sigma} U^{\mu} U^{\alpha}\Psi^{\beta}.
\end{equation}
Substituting in its corresponding Euler-Lagrange equation:
\begin{equation}
\frac{d}{dS}\frac{\partial L}{\partial {U}^{\sigma}}- \frac{\partial L}{\partial x^{\sigma}} =0,
\end{equation}
i.e.
$$
(g_{\mu \nu} \delta^{\mu}_{\sigma} \frac{D \Psi^{\nu}}{DS})_{, \rho}U^{\rho}+ ( g_{\mu \nu} \Gamma^{\nu}_{\alpha \beta} \delta^{\beta}_{\sigma}U^{\mu})_{\rho} U^{\rho}- (g_{\mu \nu , \sigma} U^{\mu} \frac{D \Psi^{\nu}}{DS}+ g_{\mu \nu} \Gamma^{\nu}_{\alpha \beta ,\sigma} U^{\mu} U^{\alpha}\Psi^{\beta}) =0
$$
The resultant equation is not tonsorially covariant unless, if one substitutes  the following quantity
 $$\frac{D U^{\alpha}}{DS} =0 ,$$  in it, and provided that the Riemann curvature is defined as
$$
R^{\alpha}_{\mu \nu \rho} ~~{\stackrel{def.}{=}}~~ \Gamma^{\alpha}_{\mu \rho , \nu} - \Gamma^{\alpha}_{\mu \nu \rho} +  \Gamma^{\epsilon}_{\mu \rho}\Gamma^{\alpha}_{\epsilon \nu} - \Gamma^{\epsilon}_{\mu \nu}\Gamma^{\alpha}_{\epsilon \rho}
$$
 then, after some manipulations we obtain:
\begin{equation}
\frac{D^{2} \Psi^{\alpha}}{DS^{2}} = R^{\alpha}_{. \beta \gamma \delta} \Psi^{\gamma} U^{\beta}U^{\delta}.
\end{equation}
The above method has been applied in different geometries than the Riemannian one e.g. non-Riemannian geometries admitting non-vanishing curvature and torsion tensors simultaneously [20-22]. This approach helps to implement the concept of geometrization  to include not only physics but also biological epidemic curves [23] as well as economic complex systems in terms of information geometry [24]. Moreover, the Bazanski Lagrangian has been modified to describe the path equation of charged object to take the following form [25];
\begin{equation}
L = g_{\alpha \beta} U^{\alpha}\frac{D \Psi^{\beta}}{DS} + \frac{e}{m} F_{\alpha \beta}U^{\alpha} \Psi^{\beta}.
\end{equation}
If we take the variation with respect to ${\Phi^{\alpha}}$ following the similar steps as in (5) we obtain
\begin{equation}
\frac{dU^{\alpha}}{d S} +\Gamma^{\alpha}_{\mu \nu}U^{\mu}U^{\nu}= \frac{e}{m}F^{\mu}_{. \nu} U^{\nu}
\end{equation}
where $F_{\mu \nu}$ is an electromagnetic tensor, $\frac{e}{m}$ the ratio between charge to mass of an object.
And, on taking the variation with respect to $U^{\alpha}$ , the equation is not tonsorially covariant till we impose the following conditions \\
 (1)$ \frac{D U^{\alpha}}{DS} = \frac{e}{m} F^{\mu}_{\nu} U^{\nu}$  \\
(2)$ F_{\mu\nu ; \rho} + F_{\rho \mu ; \nu} + F_{\mu \rho ; \nu} = 0 .$ \\
Consequently, after some manipulations, we obtain its corresponding deviation equation:
\begin{equation}
\frac{D^{2}\Psi^{\alpha}}{DS^{2}}= R^{\alpha}_{.\mu \nu\rho}U^{\mu}U^{\nu}\Psi^{\rho} +\frac{e}{m}(F^{\alpha}_{.\nu} \frac{D \Psi^{\nu}}{Ds}+F^{\alpha}_{.\nu ; \rho}U^{\nu}\Psi^{\rho}).
\end{equation}
Moreover,  for non precessing spinning objects, equations of spin and spin deviation are  obtained by applying action principle on  the following Lagrangian :
\begin{equation}
L = g_{\alpha \beta} U^{\alpha}\frac{D \Psi^{\beta}}{DS} + \frac{1}{2m} R_{\alpha \beta \gamma \sigma}U^{\alpha} \Psi^{\beta}S^{\gamma \sigma}
\end{equation}
where $S^{\mu \nu}$ is a spin tensor of a spinning object.
 By taking variation with respect to the $\Psi^{\alpha}$ we obtain
\begin{equation}
\frac{dU^{\alpha}}{dS}+\Gamma^{\alpha}_{\mu \nu}U^{\mu}U^{\nu}= \frac{1}{2m} R^{\alpha}_{. \mu \nu \rho} S^{\rho \nu} U^{\mu}
\end{equation}

And taking the variation with respect to $U^{\alpha}$ ,  and using the following condition conditions: \\
(1) $ \frac{DU^{\alpha}}{DS} = \frac{1}{2m} R^{\alpha}_{\beta \gamma \delta} S^{\gamma \delta} U^{\beta} ,$ \\
(2) $ R^{\alpha}_{\beta \gamma \delta} + R^{\alpha}_{\delta \beta \gamma} + R^{\alpha}_{\gamma \delta \beta} =0 $ \\
 to obtain its deviation equation:

\begin{equation}
\frac{D^{2}\Psi^{\alpha}}{DS^{2}}= R^{\alpha}_{.\mu \nu\rho}U^{\mu}U^{\nu}\Psi^{\rho} +\frac{1}{2m}( R^{\alpha}_{. \mu \nu \rho} S^{\nu \rho} \frac{D \Psi^{\nu}}{Ds}+
R^{\alpha}_{\mu \nu \lambda}S^{\mu \lambda}_{.; \rho}U^{\nu}\Psi^{\rho} + R^{\alpha}_{\mu \nu \lambda; \rho }S^{\nu \lambda} U^{\mu} \Psi^{\rho}),
\end{equation}
 Also, for deriving path equations for spinning charged object [27] we  take the variation with respect to $\Psi^{\alpha}$ on the following lagrangian
\begin{equation}
L= g_{\mu \nu} U^{\mu} \frac{D \Psi^{\nu}}{DS} + \frac{1}{m} (e F_{\mu \nu} + \frac{1}{2} R_{\mu \nu \rho \sigma} S^{\rho \sigma} ) U^{\nu}\Psi^{\mu}
\end{equation}
to obtain
\begin{equation}
\frac{dU^{\alpha}}{d S}+\Gamma^{\alpha}_{\mu \nu}U^{\mu}U^{\nu}=   \frac{e}{m}F^{\mu}_{. \nu} U^{\nu}+\frac{1}{2m}
R^{\alpha}_{. \mu \nu \rho} S^{\rho \nu} U^{\mu},
\end{equation}

And taking the variation with respect to $U^{\alpha}$ , providing the following consitions: \\
 (1) $ \frac{DU^{\alpha}}{DS} = \frac{e}{m} F^{\alpha}_{\beta} U^{\beta}+ \frac{1}{2m} R^{\alpha}_{\beta \gamma \delta} S^{\gamma \delta} U^{\beta} ,$ \\
(2)$ F_{\mu\nu ; \rho} + F_{\rho \mu ; \nu} + F_{\mu \rho ; \nu} = 0 .$ \\
(3) $ R^{\alpha}_{\beta \gamma \delta} + R^{\alpha}_{\delta \beta \gamma} + R^{\alpha}_{\gamma \delta \beta} =0 $ \\
we obtain its corresponding deviation equation
$$
\frac{D^{2}\Psi^{\alpha}}{Ds^{2}}= R^{\alpha}_{.\mu \nu\rho}U^{\mu}U^{\nu}\Psi^{\rho} + \frac{e}{m}( F^{\alpha_{.\nu}} \frac{D \Psi^{\nu}}{Ds} + F^{\alpha}_{.\nu ; \rho}U^{\nu}\Psi^{\rho})+ \frac{1}{2m} R^{\alpha}_{.\mu \nu\rho}U^{\mu}U^{\nu}\Psi^{\rho}
$$
\begin{equation}
~~~~~~~~~~~~~~
+
\frac{1}{2m}( R^{\alpha}_{. \mu \nu \rho} S^{\nu \rho} \frac{D \Psi^{\nu}}{Ds}+
R^{\alpha}_{\mu \nu \lambda}S^{\mu \lambda}_{.; \rho}U^{\nu}\Psi^{\rho} + R^{\alpha}_{\mu \nu \lambda; \rho }S^{\nu \lambda} U^{\mu} \Psi^{\rho})
\end{equation}

 Moreover, for a spinning object with precession we  modified Bazanski Lagrangian [29] :
\begin{equation}
L= g_{\alpha \beta} ( m U^{\alpha} + U_{\beta}\frac{D S^{\alpha \beta}}{DS}) \frac{D \Psi^{\beta}}{Ds} + \frac{1}{2} R_{\alpha \beta \gamma \delta} S^{\gamma \delta} U^{\beta} \Psi^{\alpha}
\end{equation}
to obtain equation of a spinning object by taking the variation with respect to the deviation vector $\Psi^{\alpha}$
\begin{equation}
 \frac{D}{DS}( m U^{\alpha} + U_{\beta}\frac{D S^{\alpha \beta}}{DS})= \frac{1}{2} R^{\alpha}_{. \mu \nu \rho} S^{\rho \nu} U^{\mu}
\end{equation}

And taking the variation with respect to $U^{\alpha}$ ,  and using the following condition conditions: \\
(1) $ \frac{DP^{\alpha}}{DS} = \frac{1}{2} R^{\alpha}_{\beta \gamma \delta} S^{\gamma \delta} U^{\beta} ,$ \\
(2) $ R^{\alpha}_{\beta \gamma \delta} + R^{\alpha}_{\delta \beta \gamma} + R^{\alpha}_{\gamma \delta \beta} =0 $ \\
 to obtain its deviation equation:
$$
\frac{D^{2}\Psi^{\alpha}}{Ds^{2}}=  R^{\alpha}_{.\mu
\nu\rho}U^{\mu}( m U^{\nu} + U_{\beta}\frac{D S^{\nu
\beta}}{Ds})\Psi^{\rho}+ g^{\alpha \sigma}g_{\nu \lambda}( m
U^{\lambda} + U_{\beta}\frac{D S^{\lambda \beta}}{Ds})_{; \sigma}
\frac{D \Psi^{\nu}}{Ds}
$$
\begin{equation}
~~~~~~~+ \frac{1}{2}(R^{\alpha}_{. \mu \nu \rho} S^{\nu \rho} \frac{D
\Psi^{\mu}}{Ds}+ R^{\alpha}_{\mu \nu \lambda}S^{\nu \lambda}_{.;
\rho}U^{\mu}\Psi^{\rho} + R^{\alpha}_{\mu \nu \lambda; \rho
}S^{\nu \lambda} U^{\mu} \Psi^{\rho}) .
\end{equation}
However, the spin precession equation can not be obtained from the above Lagrangian but due to some relations between $P^{\mu}$ and its unit tangent vector $U^{\alpha}$
\begin{equation}
P^{\mu} U^{\nu} = (m U^{\mu} + U_{\beta} \frac{DS^{\mu \beta}}{DS}) U^{\nu}
\end{equation}
 and
 \begin{equation}
P^{\nu} U^{\mu} = (m U^{\nu} + U_{\beta} \frac{DS^{\nu \beta}}{DS}) U^{\mu}.
\end{equation}
Thus, after some manipulations, regarding  that $ U^{\alpha}U_{\beta} = \delta^{\alpha}_{\beta} $ one obtains the equation of spin precession:
\begin{equation}
\frac{D S^{\mu \nu}}{DS} = 2(P^{\mu}U^{\nu} - P^{\nu}U^{\mu}).
\end{equation}

\subsection{Path and Path Deviation Equations in Weyl Geometry}
It is  well known that in Weyl geometry the gravitational potential tensor is associated with such a scalar field. From this perspective one can define a combined gravitation potential tensor in the following manner [30]:
  \begin{equation}
  \bar{g}^{\mu \nu} =  e^{\phi}g^{\mu \nu},
  \end{equation}
  Where $\bar g_{\mu \nu}$ is the Weyl gravitational potential, and $\phi$ a scalar field,  which  may give raise to introduce disformal transformation of any gravitational theory having two metrics defined in the following way [31]
  \begin{equation}
  \bar{g}_{\mu \nu} =[ A g_{\mu \nu} + \bar{A} \phi_{, \mu} \phi_{, \nu}]
  \end{equation}
   where $A$ and $\bar{A}$ are arbitrary constants .\\

  Thus, in this type of geometry it can be defined its corresponding  affine connection to become :
  \begin{equation}
  \bar{\Gamma}^{\alpha}_{\beta \sigma} = \Gamma^{\alpha}_{\beta \sigma} + \frac{1}{2}g^{\alpha \delta}( g_{\sigma \delta} \phi_{,\beta } +g_{\delta \beta} \phi_{, \sigma } -g_{\beta \sigma} \phi_{,\delta})
\end{equation}
 In order to obtain the geodesic equation , we introduce the following Lagrangian:
\begin{equation}
L= \bar{g}_{\mu \nu} U^{\mu} \frac{ \bar{D} \Psi^{\nu} }{\bar{D}S}.
\end{equation}
If we take the variation with respect to $\Psi^{\alpha}$, we obtain the path equation:
\begin{equation}
  \frac{dU^{\alpha}}{d S}+ \bar{\Gamma}^{\alpha}_{\beta \sigma} U^{\beta}U^{\sigma}=0,
 \end{equation}
 And its corresponding deviation equation is determined by  applying the commutation relation between the two parameters in a similar way as explained in (2.1) is obtained by taking the variation with respect to $U^{\mu}$ to become:
 \begin{equation}
 \frac{\bar{D}^{2}\Psi^{\alpha}}{\bar{D}S^{2}}= \bar{R}^{\alpha}_{.\mu \nu\rho}U^{\mu}U^{\nu}\Psi^{\rho}
 \end{equation}
 where
 $$
 \bar{R}^{\alpha}_{.\mu \nu\rho}= \bar{\Gamma}^{\alpha}_{\mu \rho ,\nu} - \bar{\Gamma}^{\alpha}_{\mu \nu ,\rho}
 + \bar{\Gamma}^{\sigma}_{\mu \rho } \bar{\Gamma}^{\alpha}_{\sigma \rho }  - \bar{\Gamma}^{\sigma}_{\mu \rho } \bar{\Gamma}^{\alpha}_{\sigma \rho }
 $$

(ii) Dixon-like Equation for  spinning charged objects of Weyl geometry: \\
Similarly ,  we can obtain the Dixon-like path equation  as defined in Weyl geometry by suggesting the following Lagrangian

 \begin{equation}
\frac{dU^{\alpha}}{dS}+ \bar{\Gamma}^{\alpha}_{\mu \nu}U^{\mu}U^{\nu}=   \frac{e}{m}F^{\mu}_{. \nu} U^{\nu}+\frac{1}{2m}
\bar{R}^{\alpha}_{. \mu \nu \rho} S^{\rho \nu} U^{\mu}
\end{equation}

and its corresponding deviation equation becomes:
$$
\frac{\bar{D}^{2}\Psi^{\alpha}}{\bar{D}S^{2}}= \bar{R}^{\alpha}_{.\mu \nu\rho}U^{\mu}U^{\nu}\Psi^{\rho} + \frac{e}{m}( F^{\alpha_{.\nu}} \frac{\bar{D} \Psi^{\nu}}{\bar{D}S} + F^{\alpha}_{.\nu ; \rho}U^{\nu}\Psi^{\rho})+ \frac{1}{2m} \bar{R}^{\alpha}_{.\mu \nu\rho}U^{\mu}U^{\nu}\Psi^{\rho}
$$
\begin{equation}
~~~~~~~~~~~~~~
+
\frac{1}{2m}( \bar{R}^{\alpha}_{. \mu \nu \rho} S^{\nu \rho} \frac{D \Psi^{\nu}}{Ds}+
\bar{R}^{\alpha}_{\mu \nu \lambda}S^{\mu \lambda}_{.; \rho}U^{\nu}\Psi^{\rho} + \bar{R}^{\alpha}_{\mu \nu \lambda; \rho }S^{\nu \lambda} U^{\mu} \Psi^{\rho})
\end{equation}

\subsection {Path and Path Deviation Equations of MOND}
In this part, it is worth mentioning the path and path deviation of modified Newtonian Dynamics (MOND)  paradigm due to its vital role in explaining the vague regions due to dark matter problem that are unknown by Newtonian/Einsteinain formulations i.e. revealing the nature of rotation curves of spiral galaxies.
 Accordingly, some authors have studied motion of a test in MOND  [32].  This has led us to apply the Bazanski method in order to obtain the path and path deviation equations for any test particle related to this paradigm by suggesting the following Lagrangian

\begin{equation}
L= g_{\mu \nu} U^{\alpha}\frac{D \Psi^{\beta}}{DS} + \frac{1}{m} \phi_{ \mu}{\Psi}^{\mu}
\end{equation}
where $\phi$ is a Newtonian potential added due to MOND . \\
Applying the same approach of the Bazanski method in (2.1) by taking the variation with respect to $\Psi^{\alpha}$  we obtain its path equations
\begin{equation}
\frac{dU^{\alpha}}{d S}+{\Gamma}^{\alpha}_{\mu \nu}U^{\mu}U^{\nu}= \frac{1}{m}g^{\alpha \mu} \phi_{ \mu}.
\end{equation}
Also, equations of geodesic deviation  are obtained by taking the variation with respect to $U^{\alpha}$ on the Lagrangian  (36) with taking into consideration that
$$
\frac{DU^{\alpha}}{DS} = \frac{1}{m}g^{\alpha \rho} \phi_{\rho},
$$
 to preserve the tensorial character of the derived equation . \\
Thus, after some manipulations we obtain the path deviation equations for an object as explained by MOND paradigm.
\begin{equation}
\frac{D^2  \Psi^{\alpha}}{DS^2} = R^{\alpha}_{\beta \gamma \delta} U^{\beta}U^{\gamma} \Psi^{\delta} +\frac{1}{m} g^{\alpha \rho}\phi_{\rho; \sigma}\Psi^{\sigma}+ \frac{1}{m}g^{\alpha \rho} \phi_{\rho}U_{\nu}\frac{D \Psi^{\nu}}{DS}.
\end{equation}

\section{Path and Path Deviation of Bi-metric Theories}

\subsection{Path Equation and Path Deviation of  Rosen's Approach}
 Equations of path and path deviation for the Rosen version of bi-metric theory of gravity is derived  due to the following Lagrangian:
\begin{equation}
L=(g_{\mu \nu}- \gamma_{\mu \nu}) U^{\mu} \frac{\nabla \Psi^{\nu}}{ \nabla S}
\end{equation}
where $ \frac{\nabla \Psi^{\nu}}{\nabla S} = {{\frac{d \Psi^\mu}{dS}} + \Delta^{\mu}_{\nu \sigma} \Psi^\nu U^\sigma } $. \\
By taking the variation with respect to $\Psi^{\alpha}$ we obtain:
\begin{equation}
\frac{\partial L}{\partial \Psi^{\sigma}} = (g_{\alpha \beta } - \gamma_{\alpha \beta}) \Gamma^{\beta}_{\mu \nu} \delta^{\nu}_{\sigma}U^{\mu},
\end{equation}

\begin{equation}
\frac{\partial L}{\partial \dot{\Psi}^{\sigma}} = (g_{\alpha \beta} - \gamma_{\alpha \beta} ) \delta^{\beta}_{\sigma}U^{\alpha},
\end{equation}
\begin{equation}
\frac{d}{dS}\frac{\partial L}{\partial \dot{\Psi}^{\sigma}}=  (g_{\alpha \sigma , \rho}- \gamma_{\alpha \sigma , \rho} ) U^{\alpha}U^{\rho} + (g_{\alpha \sigma}- \gamma_{\alpha \beta}) \frac{d U^{\alpha}}{dS}.
\end{equation}
Substituting from the above relations into the Euler-Lagrange equation:
\begin{equation}
\frac{d}{dS}\frac{\partial L}{\partial \dot{\Psi}^{\sigma}}- \frac{\partial L}{\partial \Psi^{\sigma}} =0,
\end{equation}
and taking into account the following conditions:
\begin{equation}
g_{\mu \nu ; \rho} =  0,
\end{equation}
\begin{equation}
\gamma_{\mu \nu | \rho} =  0,
\end{equation}
where $|$ is a covariant derivative with respect to $\gamma_{\mu \nu}$ .
\begin{equation}
 (g_{\rho \nu} - \gamma _{\rho \nu }) (\frac{{d} U^{\nu}}{dS}+ {\Delta}^{\nu}_{\mu \sigma }U^{\sigma} U^{\mu}) =0,
\end{equation}

Multiplying equation (46) by $g^{\alpha \rho}$ and regarding that $g^{\mu \nu } \gamma_{\mu \rho} = 0$ we obtain
\begin{equation}
 \frac{{d} U^{\alpha}}{dS}+ {\Delta}^{\alpha}_{\mu \sigma }U^{\sigma} U^{\mu} =0,
\end{equation}
 which is the same equation obtained by Rosen (1940) [4]. Accordingly, its corresponding deviation equation can be obtained by following the same technique of Bazanski approach by taking the variation with respect to $U^{\alpha}$ on (36) to obtain them - with taking into account the following condition to preserve  its tensorial character.
  $$ \frac{\nabla U^{\alpha}}{\nabla S} =0 $$
we get after some rearrangements the following geodesic deviation equation:
\begin{equation}
\frac{\nabla^2 \Psi^{\nu}}{\nabla S^2}  = ( R^{\alpha}_{\beta \gamma \sigma}- P^{\alpha}_{\beta \gamma \sigma}) U^{\beta} U^{\gamma} \Psi^{\sigma},
\end{equation}
where $P^{\alpha}_{\beta \gamma \sigma}$ is the curvature tensor obtained by the affine connection $\gamma^{\alpha}_{\beta \delta}$ [3].

Due to Rosen's approach  the  curvature tensor $P^{\alpha}_{\beta \gamma \sigma}=0$  which reduces equation (48) to become
$$
\frac{\nabla^2 \Psi^{\nu}}{\nabla S^2}   = R^{\alpha}_{\beta \gamma \sigma} U^{\beta} U^{\gamma} \Psi^{\sigma}.
$$

Also, for charged objects in bi-metric theory of gravity, Falik and Rosen [33] obtained their corresponding  field  equations , which led us   to introduces the following Lagrangian to obtain the corresponding path and path deviation equation :
\begin{equation}
L= (g_{\mu \nu}- \gamma_{\mu \nu}) U^{\mu} \frac{\nabla \Psi^{\nu}}{\nabla S} + \frac{e}{m} F_{\mu \nu}  U^{\mu} \Psi^{\nu}
\end{equation}

to give
\begin{equation}
\frac{\nabla U^{\alpha}}{\nabla S } = \frac{e}{m}F^{\alpha}_{. \nu} U^{\nu} .
\end{equation}

And its corresponding deviation equation becomes:
\begin{equation}
\frac{\nabla^{2}\Psi^{\alpha}}{\nabla S^{2}}= R^{\alpha}_{.\mu \nu \rho}U^{\mu}U^{\nu}\Psi^{\rho} + \frac{e}{m}F^{\alpha}_{. \nu} \frac{\nabla \Psi^{\nu}}{\nabla S}+ \frac{e}{m}(F^{\alpha}_{. \nu ; \rho}-F^{\alpha}_{. \nu | \rho})U^{\nu}\Psi^{\rho}
\end{equation}

Moreover, Avakian et al.[16] studied the field equations of a spinning body in the presence of bi-metric theory.
 Accordingly, we can apply the same procedure as mentioned in (2.1) on the  following Lagrangian:
\begin{equation}
L = ( g_{\alpha \beta}- \gamma_{\alpha \beta} ) U^{\alpha} \frac{\nabla \Psi^{\beta}}{\nabla S} + \frac{1}{2m} ( R_{\alpha \beta \gamma \sigma}- P_{\alpha \beta \gamma \sigma} )U^{\alpha}  \Psi^{\beta}S^{\gamma \sigma}
\end{equation}
 to obtain the path equation for a spinning object in the presence of bi-metric theory :
\begin{equation}
\frac{dU^{\alpha}}{dS}+\Delta^{\alpha}_{\mu \nu}U^{\mu}U^{\nu}= \frac{1}{2 m}( R^{\alpha}_{. \mu \nu \rho} - P^{\alpha}_{. \mu \nu \rho})S^{\rho \nu} U^{\mu} U^{\nu}
\end{equation}
and its corresponding  spinning deviation equation:

$$
\frac{\nabla^{2}\Psi^{\alpha}}{\nabla S^{2}}= ( R^{\alpha}_{.\mu \nu\rho} - P^{\alpha}_{.\mu \nu\rho} )U^{\mu}U^{\nu}\Psi^{\rho} +\frac{1}{2m}( R^{\alpha}_{. \mu \nu \rho} - P^{\alpha}_{. \mu \nu \rho} ) S^{\nu \rho} \frac{\nabla \Psi^{\nu}}{\nabla S}
$$
\begin{equation}
~~~~~~~~~~~~~~~~~~~~~~~~~~+ (R^{\alpha}_{\mu \nu \lambda}S^{\mu \lambda}_{.; \rho}- P^{\alpha}_{\mu \nu \lambda}S^{\mu \lambda}_{.| \rho})U^{\nu}\Psi^{\rho}+ (R^{\alpha}_{\mu \nu \lambda; \rho } - P^{\alpha}_{\mu \nu \lambda | \rho })S^{\nu \lambda} U^{\mu} \Psi^{\rho}
\end{equation}

Thus, if we take into consideration that $P^{\alpha}_{\beta \gamma \delta} =0$, the path equation becomes:
\begin{equation}
\frac{dU^{\alpha}}{dS}+\Delta^{\alpha}_{\mu \nu}U^{\mu}U^{\nu}= \frac{1}{2 m} R^{\alpha}_{. \mu \nu \rho} S^{\rho \nu} U^{\mu} U^{\nu}
\end{equation}
and its corresponding deviation equation becomes:

\begin{equation}
\frac{\nabla^{2}\Psi^{\alpha}}{\nabla S^{2}}=  R^{\alpha}_{.\mu \nu\rho}U^{\mu}U^{\nu}\Psi^{\rho}  +\frac{1}{2m} R^{\alpha}_{. \mu \nu \rho}  S^{\nu \rho} \frac{D \Psi^{\nu}}{Ds}+ \frac{1}{2 m}R^{\alpha}_{\mu \nu \lambda}(S^{\mu \lambda}_{.; \rho}-S^{\mu \lambda}_{.| \rho})U^{\nu}\Psi^{\rho} + \frac{1}{2 m}R^{\alpha}_{\mu \nu \lambda; \rho }S^{\nu \lambda} U^{\mu} \Psi^{\rho}
\end{equation}

\subsection{Path  and Path Deviation Equations of Moffat's Approach}
Moffat [6] presented the  framework of VSL satisfying bimetric theory and its causality to reveal the problem of dark energy due to VSL by introducing such a metric in the following way.
\begin{equation}
\hat{g}_{\mu \nu} = g_{\mu \nu}  + B \partial_{\mu}\phi \partial_{\nu}\phi
\end{equation}
where  $ \hat{g}_{\mu \nu}$ defines a specific matter metric tensor of a given matter field,  $B$ is an arbitrary constant has a dimension of ${[length]}^2$ and chosen to be positive   and $\phi$ is a bi-scalar field  . The inverse metrics $g^{\mu \nu}$ $ \hat g^{\mu \nu}$ satisfy
\begin{equation}
g^{\mu \nu}g_{\mu  \rho} =\delta^{\nu}_{\rho}
\end{equation}
$$
\hat{g}_{\mu \nu} = g_{\mu \nu}  + B \partial_{\mu} \phi \partial_{\nu} \phi
$$
\begin{equation}
\hat{g}^{\mu \nu}\hat{g}_{\mu \rho} =\delta^{\nu}_{\rho}
\end{equation}
Yet, the modification processes to control the casual propagation of the bi-scalar field led to redefine (57) to become:
\begin{equation}
\hat{g}^{\mu \nu} = g^{\mu \nu}  + \frac{B}{K} \nabla_{\mu} \phi \nabla_{\nu} \phi + KB \sqrt{T_{\mu \nu}},
\end{equation}
where $K$ is an arbitrary constant and $T_{\mu \nu}$ is a given energy-momentum tensor to control the causal propagation of the biscalar field [9]. \\
Consequently, we suggest the corresponding geodesic and geodesic equation owing to Moffat's description of bimetric theory of gravity to be obtained  by taking the variation with respect to $\Psi^{\nu}$ on the following Lagrangian
 \begin{equation}
L= \hat{g}_{\mu \nu} U^{\mu} \frac{\hat{D} \Psi^{\nu}}{\hat{D}S}
\end{equation}
to obtain its geodesic equation
\begin{equation}
 \frac{{d} U^{\nu}}{dS}+ {\hat{\Gamma}}^{\nu}_{\mu \rho }U^{\rho} U^{\mu} =0,
\end{equation}
where

\begin{equation}
\hat{\Gamma}^{\nu}_{\mu \rho }= \frac{1}{2}\hat{g}^{\sigma \nu} (\hat{g}_{\rho \sigma, \mu} + \hat{g}_{\mu \sigma, \rho} - \hat{g}_{\rho \mu , \sigma}).
\end{equation}
While, taking the variation  with respect to $U^{\alpha}$ , providing that $\frac{\hat{D} U^{\nu}}{\hat{D}S} =0$  to obtain its corresponding geodesic deviation equation:
$$
\frac{\hat{D}^2 \Psi^{\nu}}{\hat{D}S^2} = \hat{R}^{\alpha}_{\beta \gamma \delta } U^{\gamma} U^{\beta} \Psi^{\delta}
$$
where
$$
\hat{R}^{\alpha}_{\beta \gamma \delta }= \hat{\Gamma}^{\alpha}_{\beta \delta , \gamma }- \hat{\Gamma}^{\alpha}_{\beta \gamma , \delta } +\hat{\Gamma}^{\nu}_{\beta \delta }\hat{\Gamma}^{\alpha}_{\nu \gamma }-\hat{\Gamma}^{\nu}_{\beta \gamma }\hat{\Gamma}^{\alpha}_{\nu \delta }.
$$

\subsection{Path and Path Deviation Equations of BIMOND Type Theories}

In this section, we present the corresponding path and path deviation equation for  test particles or  spinning objects in the presence of BIMOND theories. Accordingly , it is worth mentioning at the beginning the above corresponding paths and there deviation equation in MOND paradigm to be extended in case of its BIMOND version its corresponding path equation  becomes:
$$
L= \hat{g}_{\mu \nu} \frac{\nabla \Psi^{\nu}}{\nabla S} + (\phi_{\alpha}- \hat{\phi}_{\alpha})\Psi^{\alpha}
$$
where: where $\hat\phi$ is an associated potential related to gravitational potential $\gamma_{\mu \nu}$ and
$$
\frac{ \nabla \Psi^{\alpha}}{\nabla S} = \frac{d\Psi^{\alpha}}{dS}+ ( \Delta^{\alpha}_{\beta \gamma}  ) \Psi^{\beta}U^{\gamma}
$$
In case of BIMOND , Milgram [14] introduced the relationship between the two affine connections as defined by $g_{\mu \nu}$ and $ \gamma_{\mu \nu}$
to become:
$$
\Delta^{\alpha}_{\beta \rho} = \Gamma^{\alpha}_{\beta \rho}- \bar\Gamma^{\alpha}_{\beta \rho},
$$
such that,
$$
g_{\mu \nu ; \rho} = g_{\delta \nu} \Delta^{\delta}_{\mu \rho}  +  g_{\delta \mu} \Delta^{\delta}_{\nu \rho}
$$
and
$$
\gamma_{\mu \nu | \rho} = - \gamma_{\delta \nu} \Delta^{\delta}_{\mu \rho}  -  \gamma_{\delta \mu} \Delta^{\delta}_{\nu \rho}.
$$

By taking the variation with respect to $\Psi^{\alpha}$ we obtain its corresponding path equation
$$
\frac{d U^{\alpha}}{d S} + ( \Delta^{\alpha}_{\beta \gamma} ) U^{\beta}U^{\gamma} = \frac{1}{m}g^{\alpha \mu}(\phi_{\mu}- \hat{\phi}_{\mu}).
$$
While if we take the variation with respect to $U^{\alpha}$ on the above Lagrangian we  its corresponding deviation equation with taking into considerations the previous steps of sec(2.1) to preserve its tensorial character.
$$
\frac{\nabla^2\Psi^{\alpha}}{\nabla S^2} = {( R^{\alpha}_{\beta \gamma \delta} - {\bar R}^{\alpha}_{\beta \gamma \delta})} \Psi^{\gamma} U^{\delta} U^{\beta} + \frac{1}{m}g^{\alpha \mu}(\phi_{\mu ; \rho}- \hat{\phi}_{\mu | \rho}) \Psi^{\rho}.
$$

 Also, in case of spinning object we introduce the following Lagrangian
 $$
L= \hat{g}_{\mu \nu} \frac{\nabla \Psi^{\nu}}{\nabla S} + \frac{1}{m}( \phi_{\alpha}- \hat{\phi}_{\alpha})\Psi^{\alpha} + \frac{1}{2m} ( R_{\alpha \beta \gamma \delta} - {\bar R}_{\alpha \beta \gamma \delta}) S^{\gamma \delta} U^{\beta}  \Psi^{\alpha} .
$$
If we take the variation with respect to $\Psi^{\alpha}$ on the above Lagrangian, its corresponding spinning equation becomes:
\begin{equation}
 \frac{ \nabla U^{\alpha}}{\nabla S}= \frac{1}{2m} ( R^{\alpha}_{. \mu \nu \rho}- \bar{R}^{\alpha}_{. \mu \nu \rho}) S^{\rho \nu} U^{\mu} + \frac{1}{m}g^{\alpha \mu}(\phi_{\mu}- \hat{\phi}_{\mu})
\end{equation}
and its spin deviation equation can be obtained by taking the variation with respect to $U^{\alpha}$ , with taking into consideration the steps of (2.1) to preserve its tensorial character we get:
$$
\frac{\nabla^{2}\Psi^{\alpha}}{\nabla S^{2}}= R^{\alpha}_{.\mu \nu\rho}U^{\mu}U^{\nu}\Psi^{\rho} + \frac{1}{2m} ( R^{\alpha}_{. \mu \nu \rho}- \bar{R}^{\alpha}_{. \mu \nu \rho}) + \frac{1}{m}g^{\alpha \mu}(\phi_{\mu ; \rho}- \hat{\phi}_{\mu | \rho}) \Psi^{\rho}
$$
\begin{equation}
~~~~~~~~~~~~~~ + \frac{1}{2m} R^{\alpha}_{. \mu \nu \rho} S^{\nu \rho} \frac{\nabla \Psi^{\nu}}{\nabla S}+
( R^{\alpha}_{. \mu \nu \rho} S^{\mu \lambda}_{.; \rho}U^{\nu}\Psi^{\rho}- \bar{R}^{\alpha}_{. \mu \nu \rho}S^{\mu \lambda}_{.| \rho}U^{\nu}\Psi^{\rho}) + ( R^{\alpha}_{\mu \nu \lambda; \rho} - \bar{R}^{\alpha}_{\mu \nu \lambda | \rho} ) S^{\nu \lambda} U^{\mu} \Psi^{\rho}
\end{equation}
\subsection{Path Equations and Path Deviation of Bi-gravity Type Theories}
Recently, Arkami et al[34] have suggested two independent metrics to explain bi-gravity phenomenon,
$$ ds^2 = g_{\mu \nu}dx^{\mu} dx^{\nu} ,$$
 and
$$d\tau^2 = h_{\mu \nu}dx^{\mu} dx^{\nu} $$. \\
Thus,  in order   to obtain geodesic-like equations of bi-gravity  theory , we following  Lagrangian [36] :
$$
( \frac{d}{d S} \frac{\partial L}{\partial \dot{\Psi}^{\alpha}} - \frac{\partial L}{\partial \Psi^{\alpha}} ) +
( \frac{d \tau}{dS} )^2 (  \frac{d}{d \tau} \frac{\partial L}{\partial \bar{\Phi}^{\alpha}}- \frac{\partial L}{\partial \Phi^{\alpha}} ) =0
$$

to  give the same results as mentioned by Arkani et al (2014)
\begin{equation}
g_{\mu \nu} \frac{DU^{\mu}}{DS} + h_{\mu \nu}{( \frac{d \tau}{dS} )}   \frac{D V^{\mu}}{D\tau} =0
\end{equation}
where $V^{\mu} = \frac{d x^{\mu}}{d \tau}$ is an associate unit tangent vector with respect to the parameter $\tau$ and $\Phi$ is its corresponding deviation vector. \\
Applying the same technique of the Bazanski approach, we  obtain its deviation equations:
to  obtain:
\begin{equation}
 g_{\mu \alpha} {[ \frac{D^2\Psi^{\alpha}}{DS^2}+ R^{\alpha}_{\beta \delta \gamma} U^{\gamma} U^{\beta} \Psi^{\delta} ]} +
{( \frac{d \tau}{d S})}^2  \gamma_{\mu \alpha} {[ \frac{D^2 \Phi^{\alpha} }{D \tau^2}+ R^{\alpha}_{\beta \delta \gamma} V^{\gamma} V^{\beta} \Phi^{\delta} ]}  ,
=0
\end{equation}

If one considers $ \frac{d \tau}{dS} \neq 0 $, the two metrics can be related to each other by means of a  quasi-metric one [31].
\begin{equation}
\tilde{g}_{\mu \nu} = g_{\mu \nu} - h_{\mu \nu} + \alpha_{g} ( g_{\mu \nu} - U_{\mu}U_{\nu} ) + \alpha_{h} ( h_{\mu \nu} - V_{\mu}V_{\nu}).
\end{equation}
Such an assumption may give rise to define its related Lagrangian of Bazanski's flavor to describe the geodesic and geodesic deviation equation due to this version of bi-gravity theory.
\begin{equation}
L = \tilde{g}_{\alpha \beta} U^{\alpha}\frac{\tilde{D} \Psi^{\beta}}{\tilde{D}S},
\end{equation}

$$
  \tilde{\Gamma}^{\alpha}_{\beta \sigma} =  \frac{1}{2}\tilde{g}^{\alpha \delta}( \tilde{g}_{\sigma \delta ,\beta }  +\tilde{g}_{\delta \beta , \sigma } -\tilde{g}_{\beta \sigma ,\delta} )
$$
 and its corresponding Lagrangian:

 \begin{equation}
 L= \tilde{g}_{\mu \nu} U^{\mu} ( \frac{d \Psi^{\nu} }{dS} + \tilde{\Gamma}^{\nu}_{\rho \delta} \Psi^{\rho} U^{\delta} )
 \end{equation}
 Thus, equation of its path equation can be obtained by taking the variation respect to $\Psi^{\mu}$ to obtain:
 \begin{equation}
 \frac{\tilde{D}U^{\alpha}}{\tilde{D}S^{2}}= 0
\end{equation}

 while taking the variation with respect to $U^{\mu}$ , with following the same steps as mentioned in (2.1) , we obtain its corresponding path deviation equation:
 \begin{equation}
 \frac{\tilde{D}^{2}\Psi^{\alpha}}{\tilde{D}S^{2}}= \tilde{R}^{\alpha}_{.\mu \nu\rho}U^{\mu}U^{\nu}\Psi^{\rho}
 \end{equation}
 where
 $$
 \tilde{R}^{\alpha}_{.\mu \nu\rho}= \tilde{\Gamma}^{\alpha}_{\mu \rho ,\nu} - \tilde{\Gamma}^{\alpha}_{\mu \nu ,\rho}
 + \tilde{\Gamma}^{\sigma}_{\mu \rho } \tilde{\Gamma}^{\alpha}_{\sigma \rho }  - \tilde{\Gamma}^{\sigma}_{\mu \rho } \tilde{\Gamma}^{\alpha}_{\sigma \rho }
 $$
Consequently, the following path and path deviation of charged and spinning objects of constant spinning tensor are explained as follows

\begin{equation}
L = \tilde{g}_{\alpha \beta} U^{\alpha}\frac{\tilde{D} \Psi^{\beta}}{\tilde{D} S} + \frac{e}{m} F_{\alpha \beta}U^{\alpha} \Psi^{\beta}
\end{equation}

to give
\begin{equation}
\frac{\tilde{D}U^{\alpha}}{\tilde{DS}} = \frac{e}{m}F^{\mu}_{. \nu} U^{\nu}
\end{equation}

and its corresponding deviation equation becomes:
\begin{equation}
\frac{\tilde{D}^{2}\Psi^{\alpha}}{\tilde{D}S^{2}}= \tilde{R}^{\alpha}_{.\mu \nu\rho}U^{\mu}U^{\nu}\Psi^{\rho} +\frac{e}{m}(F^{\alpha_{.\nu}} \frac{\tilde{D} \Psi^{\nu}}{\tilde{D}s}+F^{\alpha}_{.\nu || \rho}U^{\nu}\Psi^{\rho}).
\end{equation}
 where $ ||$ represents the covariant  derivative with respect to affine connection $\Gamma^{\alpha}_{\beta \sigma}$.
 Also, the generalized path and path deviation equations for  spinning objects are obtained from the following Lagrangian:

\begin{equation} L= \tilde{g}_{\mu \nu} \frac{\tilde{D} \Psi^{\alpha}}{\tilde{D}S} + \frac{1}{2m}\tilde{R}_{\alpha \mu \nu \rho}S^{\nu \rho}U^{\mu}\Psi^{\alpha}    \end{equation}
By taking the variation with respect $\Psi^{\alpha}$ the to obtain its corresponding path equation:
\begin{equation}
\frac{\tilde{D}U^{\alpha}}{\tilde{D}S}= \frac{1}{2m} \tilde{R}^{\alpha}_{\beta \mu \nu}S^{\mu \nu} U^{\beta}
,
\end{equation}

and taking the variation with respect $U^{\alpha}$  to obtain  its path deviation equation:
\begin{equation}
\frac{\tilde{D}^2\Psi^{\alpha}}{\tilde{D} S^2}= \tilde{R}^{\alpha}_{\beta \gamma \delta} U^{\gamma} U^{\beta} \Psi^{\delta} + \frac{1}{2m}(\tilde{R}^{\alpha}_{\beta \mu \nu }S^{\mu \nu} U^{\beta})_{|| \rho}\Psi^{\rho} + \frac{1}{2m}\tilde{R}^{\alpha}_{\beta \mu \nu }S^{\mu \nu} U^{\beta} U_{\rho}\frac{\tilde{D} \Psi^{\rho}}{\tilde{D}S}
\end{equation}
\subsection{Generalized Path and Path Deviation Equations of Bi-metric Theories}
 Hossenfelder [35] introduced an alternative version of bi-metric theory, having two different metrics $\bf{g}$ and $\bf{h}$ of Lorentzian signature on a manifold $\bf{M}$ one is defined in tangential space TM and the other is in its co-tangential space T*M respectively. These can be regarded as two sorts of  matter and twin matter, existing individually , each of them has its own field equations as defined within Riemannian geometry.
In this part we are going to present a generalized form which can be present different types of path and path deviation which can be explained for any bi-metric theory which has two different metrics and curvatures as defined by Riemannian geometry . Their Corresponding Lagrangian can be expressed in the following way:

\begin{equation}
L= g_{\mu \nu} \Psi_{; \nu} U^{\nu} + \gamma_{\mu \nu} \Phi_{| \nu} V^{\mu},
\end{equation}

By considering $ \frac{d \tau}{ds} =0$ .
This will lead to two separate sets of path equations owing to each parameter by applying the following Bazanski-like Lagrangian:

$$  L= g_{\mu \nu}\Psi_{; \nu}U^{\nu}- \gamma_{\mu \nu}{\Phi}_{| \nu}V^{\mu}V^{\nu}$$

\begin{equation}
\frac{DU^{\alpha}}{DS}=0,
\end{equation}
and
\begin{equation}
\frac{DV^{\alpha}}{D \tau}=0
\end{equation}
and their corresponding path deviation equations:
\begin{equation}
\frac{D^2\Psi^{\alpha}}{DS^2}= R^{\alpha}_{\beta \gamma \delta} U^{\gamma} U^{\beta} \Psi^{\delta},
\end{equation}
and
\begin{equation}
\frac{D^2\Phi^{\alpha}}{D\tau^2}= S^{\alpha}_{\beta \gamma \delta} V^{\gamma} V^{\beta} \Phi^{\delta},
\end{equation}

Thus we  suggest, the corresponding lagrangian to describe two independent sets of a generalized path and path deviation equations:
\begin{equation}
 L= g_{\mu \nu}\Psi_{; \nu}U^{\nu}- \gamma_{\mu \nu}{\Phi}_{| \nu}V^{\mu}V^{\nu}+ f_{\mu}\Psi^{\mu} + \hat{f}_{\mu}\Phi^{\mu}
 \end{equation}

where, $$f_{\mu} = \frac{1}{m}(e F_{\mu \nu}+ \frac{1}{2} R_{\mu \nu \rho \sigma}S^{\rho \sigma} )U^{\nu}$$
and $$\bar{f}_{\mu} = \frac{1}{m}(e F_{\mu \nu} + \frac{1}{2} S_{\mu \nu \rho \sigma}S^{\rho \sigma} )V^{\nu}.$$
By taking the variation of $\Psi^{\alpha}$ and $\Phi^{\alpha}$ respectively. \\
Consequently, we obtain a set of path equations
\begin{equation}
\frac{DU^{\alpha}}{DS}= f^{\alpha},
\end{equation}
and
\begin{equation}
\frac{DV^{\alpha}}{D \tau}= \bar{f}^{\alpha}
\end{equation}
and taking the variation with respect to $ U^{\alpha}$ and $V^{\alpha}$ using the same procedure of sec(2.1) to preserve their tensor character.  Thus, we obtain the set of their corresponding path deviation equations:
\begin{equation}
\frac{D^2\Psi^{\alpha}}{DS^2}= R^{\alpha}_{\beta \gamma \delta} U^{\gamma} U^{\beta} \Psi^{\delta} + f^{\alpha}_{; \rho}\Psi^{\rho} + g^{\alpha \rho}f_{\rho}U_{\nu}\frac{D \Psi^{\nu}}{DS}
\end{equation}
and
\begin{equation}
\frac{D^2\Phi^{\alpha}}{D\tau^2}= S^{\alpha}_{\beta \gamma \delta} V^{\gamma} V^{\beta} \Phi^{\delta} + \bar{f}^{\alpha}_{| \rho}\Phi^{\rho} + {\gamma}^{\alpha \rho}\bar{f}_{\rho}V_{\nu}\frac{D \Psi^{\nu}}{D\tau}.
\end{equation}

\section{Discussion and Concluding Remarks}
In this study, we have obtained the corresponding equations of path and path deviation equations for test particles, charged and spinning objects -constant spinning tensor- in different versions of Bimetric theories of gravity using a modified Bazanski Lagrangian. This type of study has imposed us to determine prior to its procedure some relevant path and path deviation for different  path equations in Weyl geometry and MOND paradigm to be counted as an introductory step to visualize the different stages of path and path deviation equations that must be included before dealing with different bi-metric theories of gravity. The study may give rise to search of a possible geometry able to express bi-metric theory of gravity. It can be sought that Finslerian geometry is a good candidate to express bi-metric theory of gravity as an extension of a Riemannian geometry [36]. In the mean time, path and path deviation equations using the Bazanski Lagrangian in Finsler geometry  are in preparation [37].
 Also, the above  treatment of utilizing  a symmetric affine connection, can be extended into another version of  bi-metric theory of gravity following Einstein-Cartan geometry,as an extended approach of Drummond [38] which gives rise to different types of torsion and how does it propagate with respect to metric propagation due bimetric formalism.
 Finally, this work  will  enable us to examine, the stability of objects orbiting  very strong gravitational field by solving the spin and spin deviation equations.

\section*{Acknowledgement}
The author would like to thank Professors  T.Harko , G. De Young, M.I.Wanas , M. Abdel Megied and his colleague Dr. E. Hassan  for their remarks and comments.
\section*{References}
{[1]} Misner, C. , Thorne, K.  Wheeler, J.(1973)  {\it{Gravitation}}, San Francisco, Feeman and Comp. \\
{[2]}Papapetrou, A. (1948) Proceedings of Royal Irish Academy Section A Vol {\bf{52}}, 11. \\
{[3]}Rosen, N. (1973)  Gen. Relativ.  and Gravit., {\bf{4}},  435. \\
{[4]}Rosen, N. Ann. Physics (1974){\bf{84}},455. \\
 {[5]}Yalmoz, H. (1975) Gen. Relativ.  and Gravit., {\bf{6}},  269. \\
 {[6]}Moffat, J.W. (2002) arXiv: hep-th/0208122  \\
  {[7]}Moffat, J.W.(2002)  arXiv: quant-ph/0204151 \\
 {[8]}Moffat, J.W.(2011) 	arXiv: 1110.1330 \\
 {[9]}Moffat, J.W. (2013)   arXiv: 1306.5470  \\
{[10]} Milgram, M. (1983), Astrophys. J. {\bf{270}}, 365  \\
{[11]} Milgram, M. (2009) Phys.Rev.D80:123536,2009 arXiv: 0912.0790 \\
{[12]} Milgram, M. (2014)arXiv: 1404.7661 \\
{[13]}(Milgram , M. (2014) Phys. Rev. D 89, 024027 (2014); arXiv:1308.5388 \\
{[14]} Hassan, S.F. and Rosen, Rachel.A. (2012) arXiv 1109.3515 \\
{[15]} Aoki, K. and Maeda , K. (2014) arXiv: 1409.0202 \\
{[16]} Avakian, R.M  , Churabian, E.V. and  Grigiorian, H. A. (1988) Astronomische Nachrichten {\bf{309}},229 \\
{[17]} Israelit, M. (1976) Gen. Relativ.  and Gravit., {\bf{7}},  623. \\
{[18]} Falik, D. and  Opher, R., (1980), Mon. Not R. Astr. Soc. , {\bf{192}}, 75. \\
{[19]} Bazanski, S.L. (1989) J. Math. Phys., {\bf {30}}, 1018. \\
{[20]} Wanas, M.I., Melek, M. and Kahil, M.E.(1995) Astrophys. Space Sci., {\bf {228}}, 273. \\
{[21]} Wanas, M.I., Melek, M and Kahil, M.E. (2000) Gravitation and Cosmology, Vol4, 319. \\
{[22]} Wanas, M.I. and Kahil, M.E.(1999) Gen. Rel. Grav., {\bf {31}}, 1921. ; gr-qc/9912007\\
{[23]} Kahil, M.E. (2006), J. Math. Physics {\bf {47}},052501. \\
{[24]}  Kahil, M.E. (2011), WSEAS Transaction of Mathematics,  Vol {\bf{10}}, Issue 12, 454 \\
{[25]}   Kahil, M. E. (2014), Hyperion International Journal of Econophysics and New Economy vol {\bf{7}} Issue 1,  62 \\
{[26]}Papapetrou, A. (1951), Proceedings of Royal Society London A {\bf{209}} , 248  \\
{[27]} Dixon, W. G. Dixon, (1970)  Proc. R. Soc. London, Ser. A {\bf{314}}, 499 \\
{[28]} Corinaldesi , E. and Papapetrou, A. (1951), Proceedings of Royal Society London A {\bf{209}}, 259\\
{[29]} Pavsic, M.  and Kahil, M. E. (2012) Central European Journal of Physics,  Volume {\bf{10}}, 414; arXiv:1012.2258\\
{[30]} Romero, C. , Fonseca-Neto, J.B. ad Pucheu, M.L. (2012) arXiv 1201.1469 \\
{[31]} Bekenstein, J.D. (1992) arXiv: gr-qc/9211017 \\
{[32])Kahil, M.E. and Harko ,T.(2009) Mod.Phys.Lett.A24:667; arXiv:0809.1915 \\
 {[33]} Falik, D. and Rosen, N. (1981), Gen. Relativ.  and Gravit., {\bf{13}}, 599 \\
([34])Akrami, Y., Kovisto, T. and Solomon, A.R. {(2014)} arXiv.1404.0006 \\
 {[35]}Hossenfelder,S. (2008) arXiv: 0807.2838\\
 {[36]}Foukzoun, J. , Podosenov, S.A., Potpov, A.A.  and Menkova, E. (2010) arXiv: 1007.3290\\
{[37]}Wanas, M.I , Kahil, M.E. , Kamal, M. (2015) A Paper in preparation. \\
{[38]}Drummond, I.T.(2001) Phys.Rev.D{\bf{63}}  043503 ; arXiv:astro-ph/0008234  \\

\end{document}